\documentclass[useAMS,usenatbib]{mn2e}

\usepackage{latexsym}
\usepackage{amssymb}
\usepackage[dvips]{graphicx}
\usepackage{mathrsfs}

\newcommand{\mbh}{$M_{\rmn{BH}}$}

\newcommand{\mmbh}{M_{\rmn{BH}}}

\newcommand{\mmsun}{\rmn{M}_{\odot}}

\newcommand{\CIV}{C{\sc iv}}

\newcommand{\etal}{et al.}
\newcommand{\MgII}{Mg{\sc ii}}
\newcommand{\Hb}{H$\beta$}

\title[Downsizing of supermassive black holes (II)]{Downsizing of supermassive black holes from the SDSS quasar survey (II). Extension to $z\sim4$} 
\author[M. Labita, R. Decarli, A. Treves and R. Falomo]{M. Labita$^{1}$\thanks{E-mail:
marzia.labita@gmail.com}, R. Decarli$^{1}$, A. Treves$^{1}$ and R. Falomo$^{2}$\\
$^{1}$Department of Physics and Mathematics, University of Insubria, Via Valleggio 11, I-22100 Como, Italy\\
$^{2}$INAF, Astronomical Observatory of Padova, Vicolo dell'Osservatorio 5, I-35122 Padova, Italy}
\begin{document}
\date{Accepted ... Received ...; in original form ...}
\pagerange{\pageref{firstpage}--\pageref{lastpage}} \pubyear{0000}
\maketitle
\label{firstpage}
\begin{abstract}
Starting from  the quasar sample of the Sloan Digital Sky Survey (SDSS) for which the \CIV\ line is observed, we use an analysis scheme to derive the $z$-dependence of the maximum mass of active black holes, which overcomes the problems related to the Malmquist bias. The same procedure is applied to the low redshift sample of SDSS quasars for which \Hb\ measurements are available. Combining with the results from the previously studied \MgII\ sample, we find that the maximum mass of the quasar population increases as $(1+z)^{1.64\pm0.04}$
in the redshift range $0.1\lessapprox z\lessapprox4$, which includes the epoch of maximum quasar activity.
\end{abstract}

\section{Introduction}
The dependence of the quasar black hole mass (\mbh) on redshift is of fundamental cosmological interest, since it is a direct probe of the supermassive black hole (BH) formation and evolution and of the triggering of the active galactic nuclei (AGN) phenomenon. A full understanding of the cosmic evolution of quasar population would also give important insights into the joint formation of BHs and their host galaxies. 

The basic procedure for the determination of the BH mass of AGN is through a direct application of the virial theorem. In the hypothesis that the clouds responsible for broad lines emission move in Keplerian orbits (see e.g.~Peterson \& Wandel 2000 for supporting evidence), the mass depends on the clouds velocity, which is constrained by the line-width, and the cloud distance, that was shown to be strictly linked to the continuum intensity (e.g.~Kaspi et al.~2000; Vestergaard 2002; Kaspi et al.~2005; see also references in Section 2).

In order to explore the \mbh($z$) dependence, one should take into account the Malmquist bias, related to the minimum detectable flux, which affects all the large samples of quasars and is apparent as an increase of the average observed luminosity with $z$. As \mbh\ is related to the nuclear luminosity, a Malmquist-type bias also affects the apparent trend of quasar BH masses (see e.g.~Vestergaard et al.~2008 and Kelly, Vestergaard \& Fan 2009 for a detailed bias analysis). 

In a previous paper (Labita et al.~2009, hereafter Paper I) we proposed a procedure to trace the $z$-dependence of \mbh\, overcoming the problems related to the Malmquist bias. We considered the $\sim50000$ quasars from the Sloan Digital Sky Survey (SDSS) Data Release 5 (DR5, Schneider \etal\ 2007; see also Richards et al.~2002)  for which \MgII\ line-width (FWHM) and 3000 \AA\ continuum luminosity were measured (Shen et al.~2008).  
Our scheme  requires to fit the distribution of quasars in the luminosity--FWHM plane (see also Fine et al.~2008), using as free parameters the maximum mass, the Eddington ratio limit  and the minimum detectable luminosity. We obtained a dependence with redshift of the type $\log (M_{\rm BH}^{\rm max}/{\rm M}_{\odot})\sim 0.3z+9$ in the interval $0.35\lessapprox z \lessapprox 2.25$, which indicates that  most massive BHs  exhibit  quasar activity at smaller cosmic times than less massive ones (downsizing, e.g.~Merloni 2004  and Shankar, Weinberg \& Miralda-Escud\'e 2009 for a theoretical approach).

Here we apply the  analysis described in Paper I on the SDSS quasar sample for which \CIV\ based virial BH mass estimates can be derived (Section \ref{secprocedurec4}). 
This extends the study of the $z$-dependence of the mass of active black holes to $z\approx4$, including the redshift region where the quasar activity is maximal  ($z_{\rm peak}\sim2-3$, e.g.~ Richards et al. 2006a).
To improve the coverage of the low redshift region, which is only marginally described by \MgII\ data,
the adopted procedure is also applied to the SDSS quasars for which the virial BH mass can be derived from measurements of the \Hb\ spectral range (Section \ref{secprocedurehb}). 

The dependence on redshift of quasar BH mass from the \Hb, \MgII\ and \CIV\ analysis is derived in Section \ref{evol}, and the results are discussed in Section \ref{sd}. To the best of our knowledge, this is the first time the \mbh ($z$) dependence of active black holes is derived beyond the peak of quasar activity from a large sample of AGN, taking into account the problems related to the Malmquist bias.

Throughout this paper, we adopt a concordant cosmology with $H_0=70$ ~km~s$^{-1}$~Mpc$^{-1}$, $\Omega_m=0.3$ and $\Omega_{\Lambda}=0.7$. 

\section{The \CIV\ and \Hb\ samples}
Shen et al. (2008) studied  $\sim 80000$ quasars 
included in the SDSS DR5 quasar catalogue. For $\sim 60000$ of them, they provide rest-frame line widths and monochromatic luminosities at 5100 \AA, 3000 \AA\ and 1350 \AA\, from which it is possible to evaluate the BH mass through the virial theorem (see Shen et al. 2008 for details on calibrations and procedures of spectral measurements). 

\begin{figure}
\includegraphics[width=0.375\textwidth, angle=270]{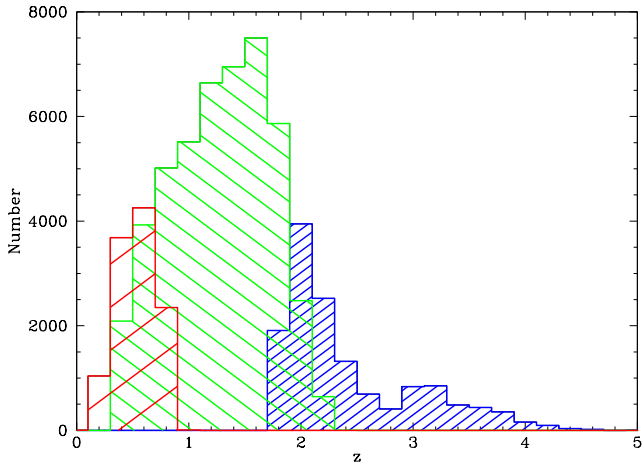}
\caption[]{Distribution in redshift of the \Hb\ sample (red thin shading), of the \MgII\ sample (green medium shading) and of the \CIV\ sample (blue thick shading).}
\label{histozhb}
\end{figure}

In Paper I, we analysed the so-called \MgII\ sample ($0.35\lessapprox z\lessapprox2.25$), containing $\sim50000$ quasars from Shen et al.~(2008) with \MgII\ measures.
Here we focus on the \CIV\ sample, consisting of the $\sim15000$ quasars from Shen et al.~(2008) for which \CIV\ line width and 1350 \AA\ monochromatic flux are available. The analysis will be limited to objects with $z<4$, since statistics are extremely poor at higher redshift. 
We also consider the low redshift ($0.1\lessapprox z\lessapprox0.9$) SDSS DR5 quasars with readily available 
\Hb\ line width and 5100 \AA\ flux ($\sim10000$ objects; hereafter, \Hb\ sample). 
All together, the \Hb, \MgII\ and \CIV\ samples cover $\sim 60000$ objects and spread over a wide redshift range, from $z\approx 0.1$ to $z\approx4$ (see Table \ref{tabsample} and Fig.~\ref{histozhb}).

\begin{table}
\centering
\caption{Redshift coverage of the \Hb, \MgII\ and \CIV\ samples.} 
\begin{tabular}{@{}ccccc@{}}
\hline
Sample& Objects &$z_{\rm min}$ & $\langle z \rangle$&$z_{\rm max}$\\ 
\hline
\Hb &11338&0.08&0.54&0.91\\ 
\MgII &46649&0.32&1.27&2.38\\ 
\CIV &14070&1.79&2.42&4.98\\ 
\hline
\label{tabsample}
\end{tabular}
\end{table}

Following Shen et al.~(2008), we assume the applicability of the virial theorem and adopt the calibration of Vestergaard \& Peterson (2006) to evaluate the BH mass for the \CIV\ sample and the calibration of McLure \& Dunlop (2004) for the \Hb\ sample:
\begin{equation}\label{formulamassa}
\log\mmbh{}=6+\log(a)+2\log({\rmn{FWHM}})+b\log{\lambda L_{\lambda}}
\end{equation} 
where $a=4.57$ and $b=0.53$ with the \CIV\ line-width and the 1350 \AA\ luminosity and $a=4.70$ and $b=0.61$ with the \Hb\ line-width and the 5100 \AA\ luminosity. Here \mbh\ is expressed in solar masses, FWHM in units of 1000 km/s and $\lambda L_{\lambda}$ in units of $10^{44}$ erg/s.
To derive the Eddington ratio ($\frac{L_{\rm bol}}{L_{\rm Edd}}$) from spectroscopic data,
we adopt the bolometric corrections by Richards et al.~(2006b). 

\section{The maximizing mass procedure}
\begin{figure}
\includegraphics[width=0.375\textwidth, angle=270]{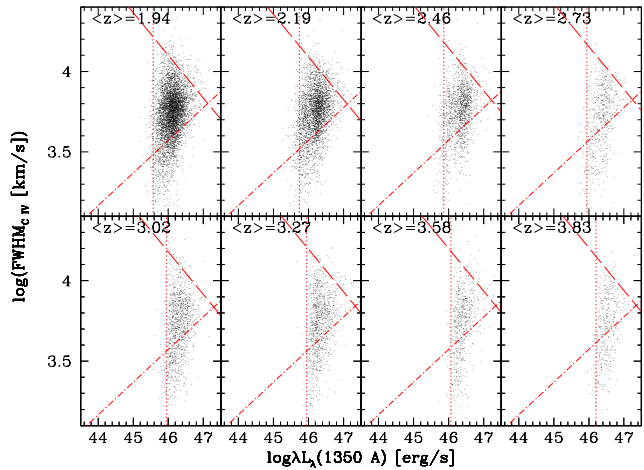}
\caption[]{The 8 panels show the \CIV\ sample in the FWHM--luminosity plane at increasing redshift. Dotted, dashed and dash-dotted lines represent the loci of constant monochromatic luminosity, constant mass and constant Eddington ratio respectively.}
\label{semplicec4}
\end{figure}

\subsection{Description of the basic procedure}
In Paper I we developed a procedure to derive the \mbh($z$) trend overcoming the problems related to the Malmquist bias, for the \MgII\ sample of SDSS quasars. The objects were divided in 8 redshift bins with constant co-moving volume. In each, we considered the objects distribution in the FWHM--luminosity plane. We aimed to construct a probability density to describe the observed distribution in terms of a minimum luminosity ($l_{\rm min}$) due to the survey flux limit, a maximum mass ($m_{\rm max}$) and a maximum Eddington ratio ($e_{\rm max}$; 
by definition, $l\equiv\log\lambda L_{\lambda}$, $m\equiv\log\frac{\mmbh}{\mmsun}$ and $e\equiv\log\frac{L_{\rm bol}}{L_{\rm Edd}}$). These three cuts form  a triangle, which describes qualitatively well the shape of the quasar distribution in the FWHM-luminosity plane (see fig.~2 of Paper I).  Then, in each redshift bin, the assumed probability density (Eq. 4 of Paper I) and the observed distribution of objects were 
discretized in 600 boxes, 
and they were compared with a best-fitting procedure in order to determine the free parameters ($l_{\rm min}$, $m_{\rm max}$ and $e_{\rm max}$ and the widths of the corresponding distributions $\sigma_l$, $\sigma_m$ and $\sigma_e$) as a function of $z$.
Errors on the best-fitting parameters were determined with Monte Carlo simulations (see Paper I for details on the procedure). 

\subsection{Analysis of the \CIV\ sample}\label{secprocedurec4}
\begin{figure*}
\begin{minipage}{\textwidth}
\centering
\includegraphics[width=0.6\textwidth, angle=270]{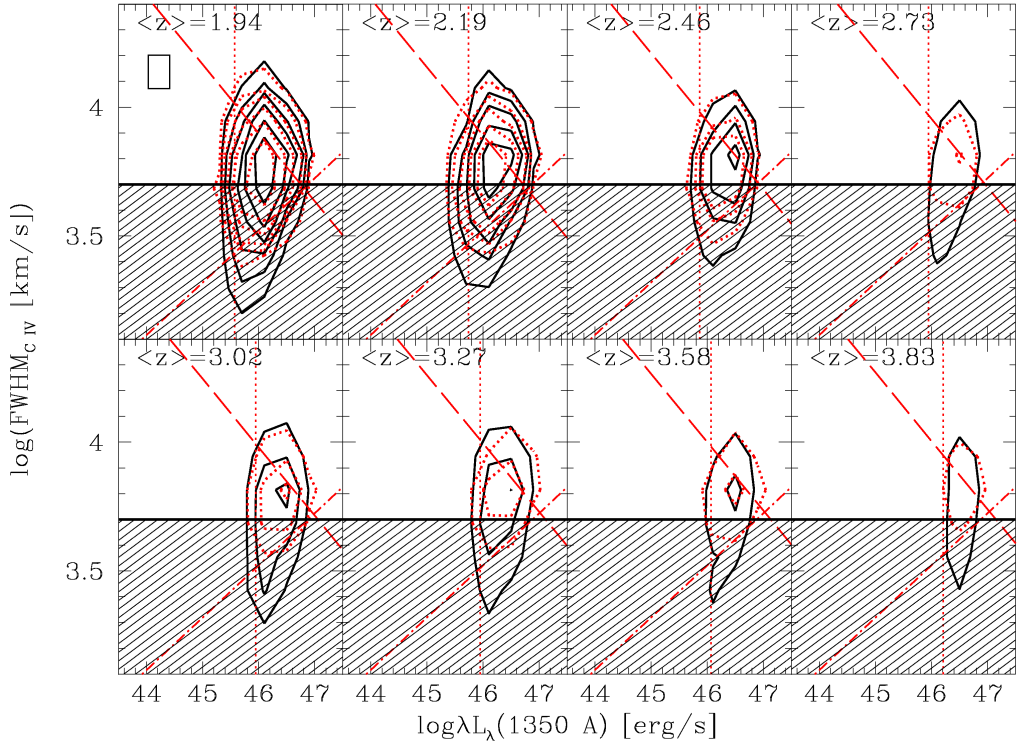}
\caption[]{The 8 panels show the \CIV\ sample in the FWHM--luminosity plane at increasing redshift: solid black contour plot (levels: 30, 80, 140, 250, 400, 750 objects per box, see text) represents the discrete observed distribution of objects. Dotted red contour plot (same levels) shows the discrete distribution of a sample of objects simulated with the Monte Carlo method, adopting the assumed $P_{\,l,\,{\rm FWHM}}(l,\,{\rm FWHM})$ probability density (Eq. 4 of Paper I) with the best-fitting parameters (see Table \ref{tabCIV}). Dotted, dashed and dash-dotted lines represent $l_{\rm min}$, $m_{\rm max}$ and $e_{\rm mIVax}$ (fixed) respectively. In each panel, the shaded area shows the region removed from the fit.  The rectangle in the first redshift bin represents the boxing adopted to discretize the distributions.}
\label{graficozCIV}
\end{minipage}
\end{figure*}

\begin{table*}
\centering
\begin{minipage}{0.67\textwidth}
\caption{Best-fitting values of minimum luminosity, maximum mass and widths of the corresponding distributions, with errors and $\chi^2_{\nu}$, for the \CIV\ sample.  In all the redshift bins, the number of degrees of freedom is $\nu=46$ (50 data-points and 4 free parameters).}
\begin{tabular}{@{}cccccccc@{}}
\hline
Bin & $<z>$ &\#(Objects)& $l_{\rm min}$ & $\sigma_l$ &  $m_{\rm max}$ & $\sigma_m$ & $\chi^2_{\nu}$\\ 
\hline 
1st &1.94&3484&{\bf 45.57}$\pm0.06$&{\bf 0.19}$\pm0.03$&{\bf  9.52}$\pm0.08$&{\bf 0.42}$\pm 0.05$& 1.55\\ 
2nd &2.20&1988&{\bf 45.74}$\pm0.07$&{\bf 0.27}$\pm0.05$&{\bf  9.50}$\pm0.10$&{\bf 0.42}$\pm 0.05$& 2.44\\ 
3rd &2.46&982&{\bf 45.86}$\pm0.10$&{\bf 0.25}$\pm0.08$&{\bf  9.61}$\pm0.07$&{\bf 0.40}$\pm 0.06$& 2.22\\ 
4th &2.73&325&{\bf 45.94}$\pm0.08$&{\bf 0.23}$\pm0.05$&{\bf  9.63}$\pm0.08$&{\bf 0.44}$\pm 0.05$& 1.33\\ 
5th &3.02&643&{\bf 45.95}$\pm0.08$&{\bf 0.24}$\pm0.06$&{\bf  9.68}$\pm0.08$&{\bf 0.43}$\pm 0.07$& 2.10\\ 
6th &3.27&605&{\bf 45.95}$\pm0.10$&{\bf 0.29}$\pm0.06$&{\bf  9.71}$\pm0.09$&{\bf 0.48}$\pm 0.08$& 1.74\\ 
7th &3.58&362&{\bf 46.06}$\pm0.09$&{\bf 0.27}$\pm0.06$&{\bf  9.72}$\pm0.11$&{\bf 0.46}$\pm 0.08$& 1.00\\ 
8th &3.83&236&{\bf 46.20}$\pm0.08$&{\bf 0.23}$\pm0.06$&{\bf  9.73}$\pm0.11$&{\bf 0.45}$\pm 0.07$& 1.32\\ 
\hline
\label{tabCIV}
\end{tabular}

{\it Note.} Data that come from a best-fitting procedure are displayed in boldface.
\end{minipage}
\end{table*}

We aim to reproduce our analysis scheme on the \CIV\ sample. A key point is that the \CIV\ emitting region is 
disc-like (e.g.~McLure \& Dunlop 2002; Labita et al.~2006, Decarli et al.~2008), as it is suggested by considerations on the broad lines shape and from the geometrical factor $f$, which links the observed FWHM of the line to the virial velocity (see McLure \& Dunlop 2002 for definition and discussion on the geometrical factor). The emission line broadening is due to Doppler effect, and it is related only to the velocity component in the line of sight direction. If the \CIV\ broad line region (BLR) is disc-like and it is observed face-on, the line width is null independently of the virial velocity. A consequence of this is that the data distribution in the FWHM--luminosity plane (Fig.~\ref{semplicec4}) is stretched towards lower values of the line width, with respect to the distribution that we would obtain if the BLR was isotropic. 

This implies that the observed distribution of objects in the FWHM--luminosity plane is not  well described by the assumed probability density (Eq. 4 of Paper 1), which is constructed under the assumption that FWHM is proportional to the BLR virial velocity. In fact, if we try to reproduce the fit procedure described in Paper I to the \CIV\ sample, we obtain that the $\chi^2_{\nu}$ values of the best fits are  much bigger than those obtained in the analysis on the \MgII\ sample. 
In particular, the description results mediocre for low values of the FWHM, while the minimum luminosity and the maximum mass sides of the triangle still describe adequately the data distribution, confirming that it appears stretched towards lower values of FWHM as expected. To rule out the possibility that this is due to bad spectral measurements, we checked the FWHM values provided by Shen et al.~(2008) for a randomly selected subsample of quasars with narrow \CIV\ lines, finding a very good agreement.

In order to overcome the problem, we restrict our fit procedure to the region of the FWHM--luminosity plane with FWHM$>$5000 km/s, which corresponds to those objects that are observed at high inclination angles ($\theta>20^{\circ}-25^{\circ}$ where $\theta$ is the angle between the line of sight and the normal to the disc plane, Decarli et al.~2008). 
Of course there will still be some orientation effects in the FWHM$>$5000 km/s sample. However, the key point is that at these line-widths the FWHM distribution predicted under isotropy assumptions is practically indistinguishable from that predicted in a disc-like BLR picture (see again Decarli et al.~2008, and in particular fig.~6 of their paper). 

Using this approach we lose the information on the maximum Eddington ratio, since the parameters $e_{\rm max}$ and $\sigma_e$ are no more constrained as they influence the shape of the probability density mostly in the lower part of the FWHM--luminosity plane. We thus fix the values of these parameters to those obtained from the \MgII\ sample ($e_{\rm max}\equiv-0.34$ and $\sigma_e\equiv0.22$). This choice is  supported by the fact that they appear to be independent of $z$ in the redshift range studied through the \MgII\ data ($0.35<z<2.25$, see Paper I). 

We then apply the fit procedure described in Paper I on the high FWHM region of the FWHM--luminosity plane with 4 free parameters ($l_{\rm min}$, $\sigma_l$, $m_{\rm max}$ and $\sigma_m$) to be constrained in all the redshift bins.  The value of $l_{\rm min}$ is not constrained by cosmology as in Paper I, because the SDSS DR5 flux limit is not homogeneous. This effect is particularly pronounced at $z\sim3$, because the selection algorithm basically targets ultraviolet excess quasars to $i=19.1$ and higher redshift ($z\gtrsim3$) quasars to $i=20.2$ (see Richards et al.~2002 and \verb+http://www.sdss.org/dr5/algorithms/target.html+ for a detailed description of the SDSS target selection). 

The expected and the observed distribution of objects are discretized in boxes with $\Delta \log$FWHM=0.13 dex and
$\Delta \log \lambda L_{\lambda}$=0.4 dex, in order to preserve sufficient statistics per box without reducing resolution significantly. The uncertainty on the best-fitting parameters induced by a different boxing is negligible with respect to the errors evalued through the Monte Carlo procedure. We also verified that the results are independent of the adopted division in redshift bins. 

Fig.~\ref{graficozCIV} shows the best-fitting Monte Carlo simulated distributions compared to the observed distributions of quasars and it is apparent that the fit is good in the allowed region of the plane. Table \ref{tabCIV} contains the best-fitting values of the free parameters, their errors and the corresponding $\chi^2_{\nu}$ values, which indicate that the quality of the fit is comparable with that of the \MgII\ sample.

The results presented here were obtained under the hypothesis that the anomalies of the distribution of quasars in the \CIV\ FWHM -- nuclear luminosity plane are basically due to a flattened geometry of the corresponding BLR. In order to test this assumption, we re-derived the probability density $P_{\,l,\,{\rm FWHM}}(l,\,{\rm FWHM})$ convolving Eq.~4 of Paper I with the orientation effects predicted under the hypothesis of a geometrically thin broad line region, in which the disc-like component of the gas velocity field accounts for the 90 per cent, and assuming that an AGN appears as a quasar if it is observed with an inclination angle $\theta<50^{\circ}$. Within this assumptions, and adopting the previously derived values of the free parameters (Table \ref{tabCIV}) the quality of the description of the observed distribution of quasars in the  \CIV\ FWHM -- nuclear luminosity plane is good at any line-width, 
indicating that the observed \CIV\ FWHM distribution is well consistent with being ascribed to orientation effects. However, the best-fitting procedure becomes impracticable with the $P_{\,l,\,{\rm FWHM}}(l,\,{\rm FWHM})$ function considered here, because it would require to introduce a large number of 
free parameters (e.g.~the fraction corresponding to the disc-like component of the gas velocity field, the maximum value of the inclination angle and their uncertainties). Therefore, the reduction of the fitting area to FWHM values greater than 5000 km/s appears to be the only viable way to proceed  with present data.

\subsection{Analysis of the \Hb\ sample}\label{secprocedurehb}

\begin{figure}
\includegraphics[width=0.375\textwidth, angle=270]{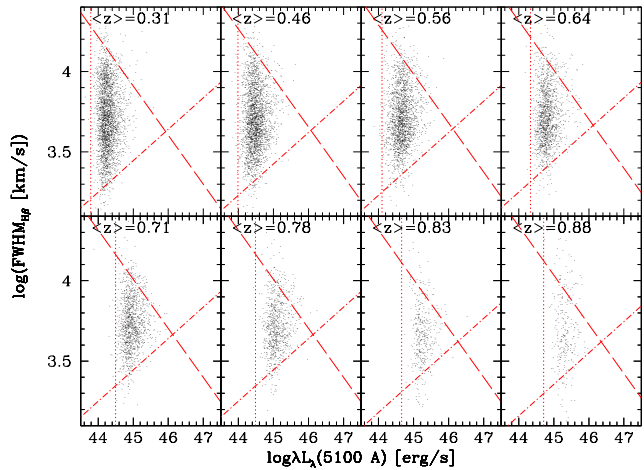}
\caption[]{The 8 panels show the \Hb\ sample in the FWHM--luminosity plane at increasing redshift. Dotted, dashed and dash-dotted lines represent the loci of constant monochromatic luminosity, constant mass and constant Eddington ratio respectively.}
\label{semplicehb}
\end{figure}

\begin{figure*}
\begin{minipage}{\textwidth}
\centering
\includegraphics[width=0.6\textwidth, angle=270]{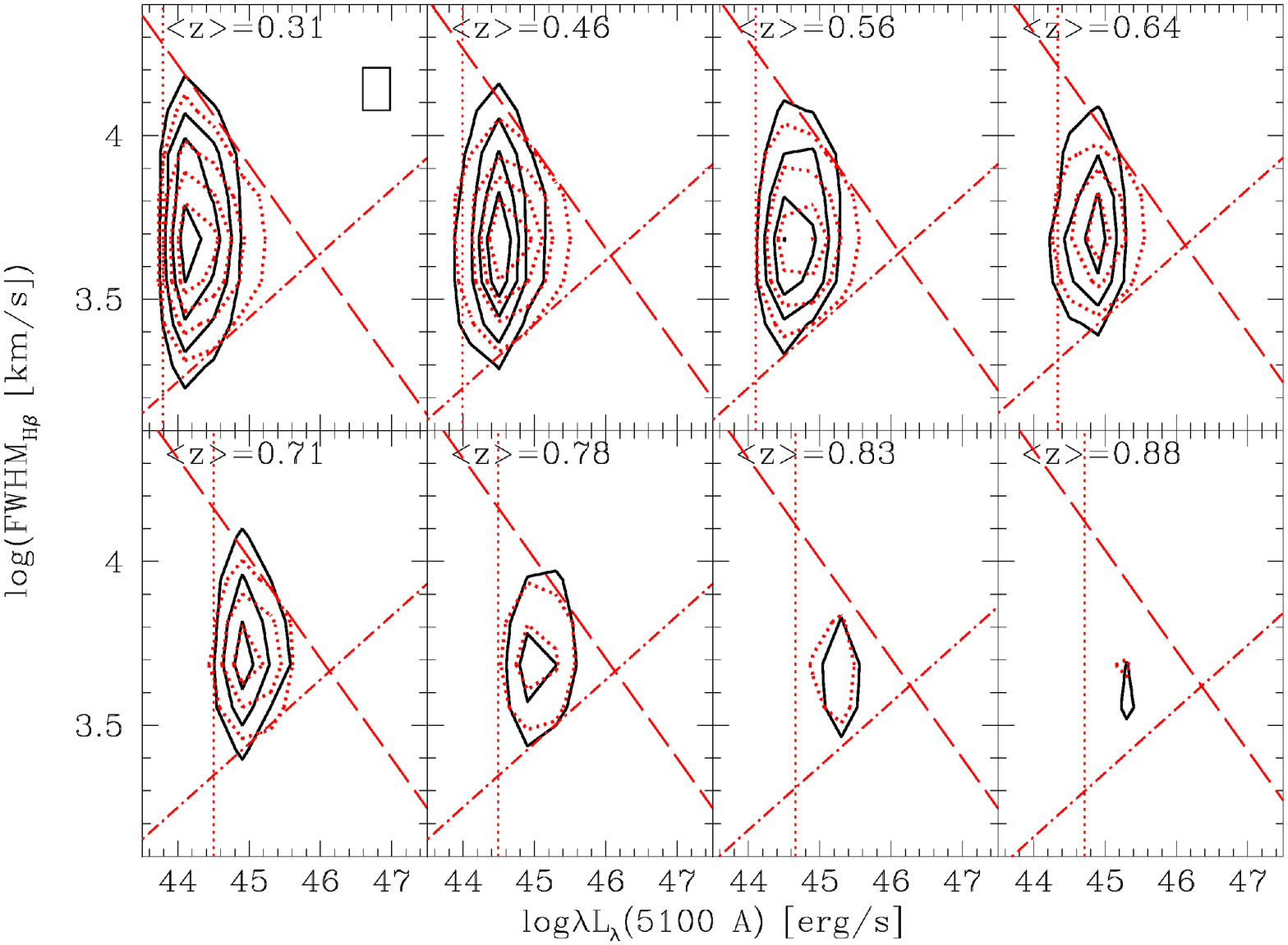}
\caption[]{The 8 panels show the \Hb\ sample in the FWHM--luminosity plane at increasing redshift: solid black contour plot (levels: 40, 110, 200, 300 objects per box, see text) represents the discrete observed distribution of objects. Dotted red contour plot (same levels) shows the discrete distribution of a sample of objects simulated with the Monte Carlo method, adopting the assumed $P_{\,l,\,{\rm FWHM}}(l,\,{\rm FWHM})$ probability density (Eq. 4 of Paper I) with the best-fitting parameters (see Table \ref{tabHB}). Dotted, dashed and dash-dotted lines represent $l_{\rm min}$, $m_{\rm max}$ and $e_{\rm max}$ respectively. The rectangle in the first redshift bin represents the boxing adopted to discretize the distributions.}
\label{graficozHB}
\end{minipage}
\end{figure*}

\begin{table*}
\centering
\begin{minipage}{0.88\textwidth}
\caption{Best-fitting values of minimum luminosity, maximum mass, maximum Eddington ratio and widths of the corresponding distributions, with errors and $\chi^2_{\nu}$, for the \Hb\ sample.  In all the redshift bins, the number of degrees of freedom is $\nu=94$ (100 data-points and 6 free parameters).}
\begin{tabular}{@{}cccccccccc@{}}
\hline
Bin & $<z>$ &\#(Objects)& $l_{\rm min}$ & $\sigma_l$ &  $m_{\rm max}$ & $\sigma_m$ & $e_{\rm max}$ & $\sigma_e$ & $\chi^2_{\nu}$ \\
\hline 
1st &0.31&2689&{\bf 43.79}$\pm0.01$&{\bf 0.10}$\pm0.01$&{\bf  9.10}$\pm0.15$&{\bf 0.35}$\pm 0.05$&{\bf -0.35}$\pm0.08$&{\bf 0.23}$\pm0.02$&13.0\\
2nd &0.46&2464&{\bf 43.99}$\pm0.02$&{\bf 0.20}$\pm0.03$&{\bf  9.20}$\pm0.15$&{\bf 0.31}$\pm 0.06$&{\bf -0.31}$\pm0.07$&{\bf 0.23}$\pm0.03$&13.9\\
3rd &0.56&2008&{\bf 44.10}$\pm0.03$&{\bf 0.24}$\pm0.03$&{\bf  9.25}$\pm0.13$&{\bf 0.35}$\pm 0.05$&{\bf -0.31}$\pm0.06$&{\bf 0.23}$\pm0.02$&6.7\\
4th &0.64&1428&{\bf 44.34}$\pm0.03$&{\bf 0.21}$\pm0.02$&{\bf  9.30}$\pm0.14$&{\bf 0.30}$\pm 0.04$&{\bf -0.34}$\pm0.07$&{\bf 0.22}$\pm0.02$&9.0\\
5th &0.71&1256&{\bf 44.50}$\pm0.03$&{\bf 0.20}$\pm0.01$&{\bf  9.30}$\pm0.10$&{\bf 0.32}$\pm 0.03$&{\bf -0.35}$\pm0.04$&{\bf 0.22}$\pm0.01$&29.6\\
6th &0.78&834&{\bf 44.49}$\pm0.02$&{\bf 0.25}$\pm0.01$&{\bf  9.30}$\pm0.08$&{\bf 0.30}$\pm 0.02$&{\bf -0.35}$\pm0.03$&{\bf 0.22}$\pm0.01$& 9.5\\
7th &0.83&392&{\bf 44.66}$\pm0.02$&{\bf 0.20}$\pm0.02$&{\bf  9.30}$\pm0.09$&{\bf 0.32}$\pm 0.03$&{\bf -0.21}$\pm0.04$&{\bf 0.22}$\pm0.01$& 13.7\\
8th &0.88&266&{\bf 44.71}$\pm0.03$&{\bf 0.21}$\pm0.02$&{\bf  9.35}$\pm0.11$&{\bf 0.30}$\pm 0.03$&{\bf -0.17}$\pm0.05$&{\bf 0.22}$\pm0.01$& 24.2\\
\hline
\label{tabHB}
\end{tabular}

{\it Note.} Data that come from a best-fitting procedure are displayed in boldface.
\end{minipage}
\end{table*}

In Fig.~\ref{semplicehb} we show the \Hb\ sample, divided in 8 redshift bins, in the FWHM--luminosity plane.
We now apply the procedure described in Paper I to this sample. Again we discretize the distributions in 100 boxes ($\Delta \log$FWHM=0.13 dex;
$\Delta \log \lambda L_{\lambda}$=0.4 dex). 
We let $l_{\rm min}$ vary as a free parameter in all the redshift bins. 

In Fig.~\ref{graficozHB} the observed distributions of objects at each redshift are compared to the distribution simulated with our procedure, and Table \ref{tabCIV} contains the best-fitting values of the free parameters and the corresponding $\chi^2_{\nu}$ values, that indicate that the quality of the fit is 
lower than that of the \MgII\ sample. In the following,  when comparing the \Hb\ maximum mass estimates with those based on \MgII\ and \CIV\ data, this will be taken into account as the errors derived for the \Hb\ best-fitting determinations are somewhat larger.

Even if there are indications that the BLR clouds where the \Hb\ line is produced are substantially isotropic, as suggested by consideration on the emission line shape (on this point see also the thick disc model in Collin et al.~2006; Decarli et al.~2008), we test whether the reduction of the fitted region to high FWHM values (see Section \ref{secprocedurec4}) would allow a better agreement between the simulated and the observed distribution: It is not the case, confirming that the explanation of the moderate quality fit of the \Hb\ data is probably uncorrelated with the inclination angles. 

We also test whether the lower quality of the results is due to the fact that the \Hb\ sample is limited to low-$z$ values and covers a small redshift window. 
In order to rule out also this possibility, 
we consider a subsample of the \MgII\ dataset assuming $z<0.91$, so that a similar redshift range is covered by both the samples (see Table \ref{tabsample}). 
The \MgII\ matched sample is split in 8 redshift bins, accordingly to the division defined for the \Hb\ dataset. The application of the maximizing mass procedure to this subsample 
gives 
$\chi^2_{\nu}$ values smaller than 5 in each bin, consistently with those obtained for the \MgII\ sample (Table 1 of Paper I) and significantly smaller than those derived for the \Hb\ sample, in the same redshift window. Since the objects included in the \MgII\ matched sample coincide with those appearing in the \Hb\ one, apart for the first redshift bin which results less populated, this test indicates that
the anomalies of the fit procedure reside right in the \Hb-based BH mass estimates (difficulties in spectral measurements or exoticism of the BLR geometry). 

\begin{figure}
\includegraphics[width=0.375\textwidth, angle=270]{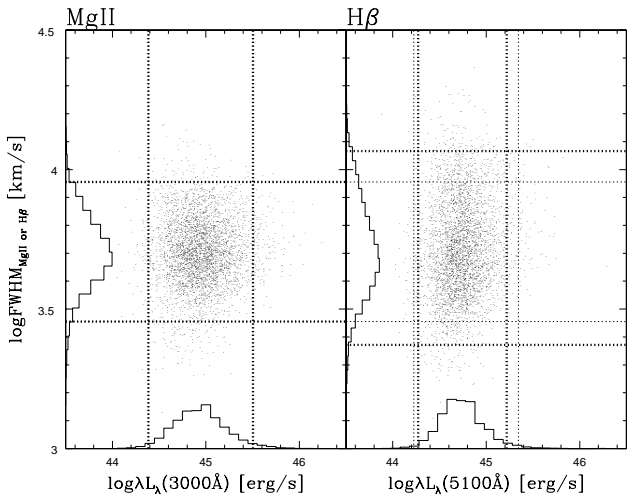}
\caption[]{The objects common to the \MgII\ and \Hb\ sample with $0.5\leq z\leq0.7$ are plotted in the FWHM--luminosity plane. The FWHM and luminosity distributions are also shown (arbitrary units) for both \MgII\ and \Hb\ data. Vertical thick lines correspond to the $\pm2\sigma$ spread in FWHM and horizontal thick lines indicate the $\pm2\sigma$ dispersion in luminosity. 
In right panel, thin lines trace the FWHM and luminosity dispersion of the \MgII\ panel, assuming a spectral index $\alpha\sim-0.3$ (Labita et al.~2008) to derive $\log\lambda L_{\lambda}(5100 {\rm \AA})$ from $\log\lambda L_{\lambda}(3000 {\rm \AA})$.}
\label{fwlum}
\end{figure}

Comparing the \Hb\ to the \MgII\ FWHM distributions for the objects for which both lines are available, we see that both the high and the low velocity tails of the \Hb\ FWHM distribution are substantially enhanced with respect to the \MgII\ one, although the widths of the two emission lines have comparable mean values and correlate reasonably (e.g.~McLure \& Jarvis 2002, Shen et al.~2008). The standard deviation of the \Hb\ FWHMs (
2525 km/s) is considerably higher than that of the \MgII\ values (
1698 km/s). On the other hand, for the same objects, the distribution of $\lambda L_{\lambda}(3000 {\rm \AA})$ is well consistent with that of $\lambda L_{\lambda}(5100 {\rm \AA})$, assuming a spectral index $\alpha\sim-0.3$ (Labita et al.~2008) to compare the luminosities at different wavelengths (see Fig.~\ref{fwlum}). This suggests that the anomalies of our results from the \Hb\ sample reside in particular in the odd distribution of the \Hb\ line-widths.

We then check whether the FWHM values provided by Shen et al.~(2008), which come from an automatized routine, are subject to systematic errors and verified the \Hb\ line width measurements for a subsample of randomly selected objects. In general we find a good agreement, apart from some cases in which the FWHM measured by Shen et al.~(2008) is very high ($\sim12000$ km/s), while we obtained considerably smaller values. This is probably due to difficulties in the deconvolution of the broad and narrow line components, which is a problem peculiar to the \Hb\ line. However, this problem affects only a small number of objects, and the dispersion of the \Hb\ line widths is likely really higher than that of the \MgII\ FWHMs. 
This would imply that the \Hb\ and \MgII\ emitting regions are intrinsically different. Note that this scenario is hardly consistent with the widely accepted assumption that the \Hb\ and \MgII\ lines are produced in the same region, given that they have similar ionizing potentials  (e.g. Verner et al. 1996), unless we suppose that the \Hb\ line is produced in clouds with  lower column densities with respect to the \MgII\ line. This would imply that the effects of radiation pressure are stronger and that the \Hb\ emitting clouds may be gravitationally unbounded (Marconi et al.~2008), accounting for non virial motion observed from the \Hb\ emission.

To explain the larger \Hb\ FWHM standard deviation with respect to the \MgII\ one,
a reasonable possibility is then to consider that velocities due to winds,  turbulent motion or inflowing gas, comparable to the Keplerian ones, may be present in the \Hb\ emitting region (see e.g.~Ruff 2008; Marconi et al.~2008; Gaskell, Goosmann \& Klimek 2008 for models and observational evidences). 
This would also account for the mediocre description of the \Hb\ sample distribution in the FWHM--luminosity plane in terms of Eq.~4 of Paper I, as the assumption of the applicability of the virial theorem is a fundamental point of our procedure.

\begin{figure*}
\begin{minipage}{\textwidth}
\centering
\includegraphics[width=0.8\textwidth, angle=0]{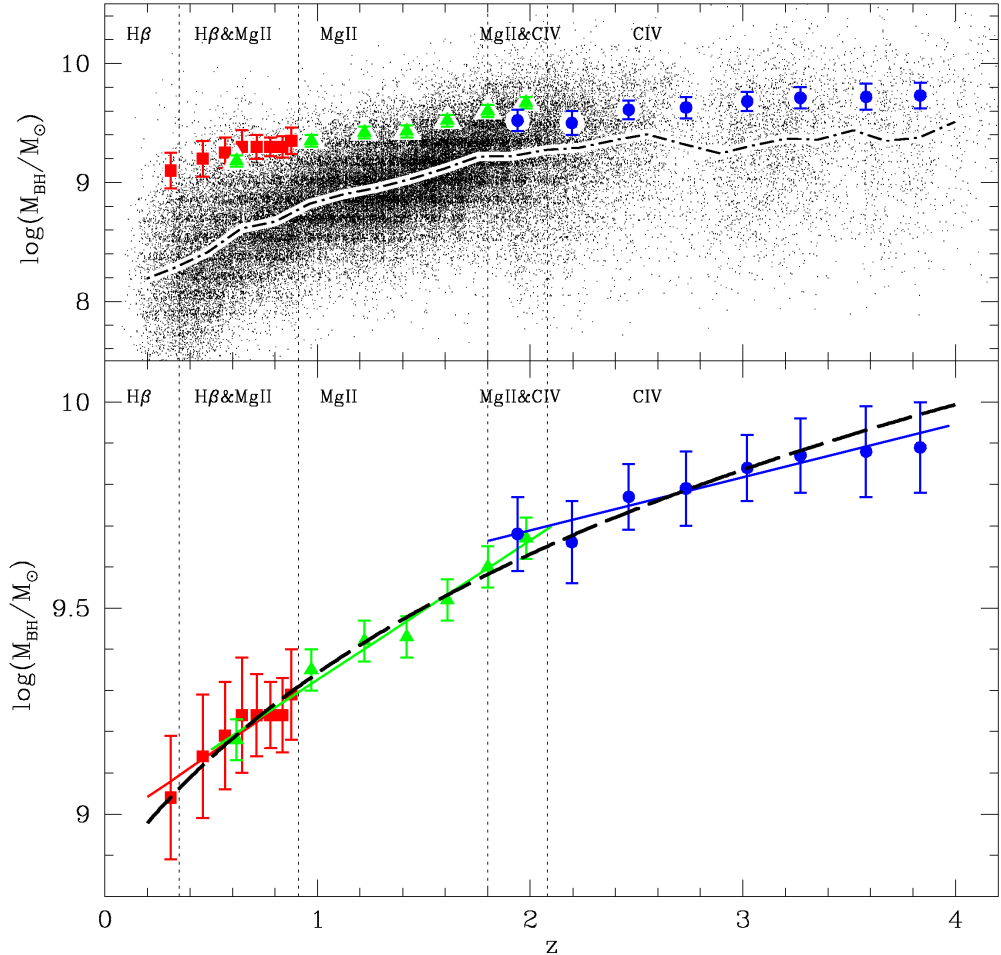}
\caption[]{{\it Upper panel:} small dots are the the virial BH masses from Shen et al.~(2008); the dash-dotted line reports the corresponding mean values. Red squares are our estimates of $\log \frac{M_{\rm BH}^{\rm max}}{{\rm M}_{\odot}}$ from the \Hb\ sample (Table \ref{tabHB}, column $m_{\rm max}$); green triangles are the same estimates from the \MgII\ sample (table 1 of Paper I, column $m_{\rm max}$) and blue circles are from the \CIV\ sample (Table \ref{tabCIV}, column $m_{\rm max}$). Vertical dotted lines indicate the redshift regions in which we derived reliable \Hb , \MgII\ and \CIV\ $m_{\rm max}$ estimates. 
{\it Lower panel:} same legend as the upper panel, but the maximum mass estimates from \Hb\ and \CIV\ were matched to the \MgII\ estimates (see text). The corresponding red, green and blue solid lines are the best linear fit to the data from the \Hb, \MgII\ and \CIV\ samples (Table \ref{chi2vfit}). The black dashed line is the best fit to all the data-points (Eq.~\ref{mmax_zall}). }
\label{mezall}
\end{minipage}
\end{figure*}

\section{Quasar BH mass dependence on redshift}\label{evol}
\subsection{Combining the \Hb , \MgII\ and \CIV\ samples}
The maximum masses in various redshift bins calculated from the three samples are reported in Fig.~\ref{mezall} (upper panel). 
The $M_{\rm BH}^{\rm max}(z)$ trend appears remarkably smooth, apart from modest offsets between the  BH mass estimates derived from different lines.

The origin of the two offsets, which are not present in the average values per redshift bin of the BH masses from Shen et al.~(2008), is probably different.  In the case of the \Hb- and \MgII-based mass estimates (hereafter, \mbh$_{{\rm , H}\beta}$ and \mbh$_{\rm , MgII}$), it is principally due to the fact that our points are representative of the maximum mass and that the \Hb\ and \MgII\ FWHM distributions are different, in the sense 
that the \Hb\ FWHM standard deviation is larger than the \MgII\ one (see Section \ref{secprocedurehb} and Fig.~\ref{fwlum}). 
This implies that, for objects with large line-widths (and then likely high masses), \mbh$_{{\rm , H}\beta}$ is on average larger than \mbh$_{\rm , MgII}$. From the data by Shen et al.~(2008), in fact, we obtain that the difference of the BH mass mean values of the objects with both \Hb\ and \MgII\ estimates is negligible ($\Delta \mmbh=0.03$ dex), while it is $\Delta \mmbh=0.13$ dex in the subsample with both \mbh$_{{\rm , H}\beta}$ and \mbh$_{\rm , MgII}$ greater than $10^9$M$_{\odot}$. 

On the other hand, the offset between the \MgII- and \CIV-based BH masses is connected to the choice of the FWHM threshold adopted in the analysis of the \CIV\ sample (Section \ref{secprocedurec4}). Assuming a higher FWHM limit, the resulting $m_{\rm max}$ estimates would be larger (for instance, it is possible to cancel the offset with a threshold at 7000 km/s), but 
the trend in redshift remains unchanged,  although statistics are significantly impoverished.  To further test the effects of an FWHM cut on the derived redshift dependence of the maximum black hole masses, we reproduced the entire procedure to the \MgII\ sample studied in Paper I, reducing the analysis to objects with \MgII\ line-widths larger than 5000 km/s. We verified that the derived slope of the $\log M_{\rm BH}^{\rm max}(z)$ dependence 
is consistent with that of Eq.~11 of Paper I within $1\sigma$, thus confirming 
that it is independent of the FWHM cut. 

For these reasons, we apply a rigid shift to the \Hb\ and \CIV\ $m_{\rm max}$ values,  imposing that they match the \MgII-based estimates in the common redshift regions. The shift amounts to -0.06 dex for the \Hb\ data-points and to +0.16 dex for the \CIV\ maximum mass estimates. 

\begin{table}
\centering
\caption{Best linear fit to the maximum mass  as a function of redshift for the \Hb\ sample, the \MgII\ sample and the \CIV\ sample. The corresponding $\chi^2_{\nu}$ values and the probability of getting worse results is also given.} 
\begin{tabular}{@{}cccccc@{}}
\hline
 &\multicolumn{2}{c}{$\log \frac{M_{\rm BH}^{\rm max}}{{\rm M}_{\odot}}=\alpha z+\beta$}\\

Sample & $\alpha$ & $\beta$  &$\nu$& $\chi^2_{\nu}$&$P(\chi'^2_{\nu}>\chi^2_{\nu})$\\
\hline
\Hb   &$0.35\pm0.05$  &$8.97\pm0.04$  &6&0.05&$>99$\%\\ 
\MgII &$0.34\pm0.02$  &$8.99\pm0.03$  &5&0.29&$\sim92$\%\\
\CIV  &$0.13\pm0.02$  &$9.45\pm0.05$  &6&0.10&$\sim99$\%\\ 
\hline
\label{chi2vfit}
\end{tabular}
\end{table}

\subsection{\mbh($z$) relationship from $z\sim0.1$ to $z\sim4$}
The shifted values of $m_{\rm max}$ (see previous section) are plotted in Fig.~\ref{mezall} (lower panel) as a function of redshift. 
In Table \ref{chi2vfit} we report the best linear weighted fits from the \Hb\ sample (Section \ref{secprocedurehb}), the \MgII\ sample (Paper I) and the \CIV\ sample (Section \ref{secprocedurec4}). 
It is worth to notice that the $\chi^2_{\nu}$ of the three best-fitting lines are all smaller than 1 and this may indicate that the Monte Carlo procedure overestimates the errors on the $m_{\rm max}$ values. 
Rather, the uncertainties that we propose are representative of the quality of the description of the observed distribution of quasars in the FWHM--luminosity plane.

As the slopes of the log(\mbh)~--~$z$ dependence 
decrease at increasing redshift, i.e.~from the \Hb\ to the \MgII\ and to the \CIV\ sample, 
we choose to describe the overall trend of the maximum quasar BH mass estimates with a power law. The best fit is:
\begin{equation}\label{mmax_zall}
\frac{M_{\rm BH}^{\rm max}}{{\rm M}_{\odot}}\propto(1+z)^{1.64\pm0.04}
\end{equation}
We stress that 
the importance of Eq. \ref{mmax_zall} is limited to the resulting trend in redshift rather than to the normalization factor, which comes from a number of assumptions on the definition of `maximum mass'.  

The overall best-fitting function is practically indistinguishable from Eq. 11 of Paper I in the \MgII\ redshift range (see Figure \ref{mezall}, lower panel). We thus confirm and extend the results of Paper I  on a wider redshift range.

Assuming that the shapes of \mbh\ and Eddington ratio distributions do not change appreciably with redshift, which is suggested by the fact that the value of $\sigma_m$  are practically independent of $z$ for each virial estimator (see Tables \ref{tabHB}, \ref{tabCIV} and table 1 of Paper I), Eq. \ref{mmax_zall} also describes the trend in redshift of the mean quasar BH mass. We conclude that quasar samples at higher $z$ are increasingly dominated by higher mass BHs, i.e. the 
lower is the mass the longer is the cosmic age at which quasars exhibit their activity.  More precisely, 
we find evidence that the most massive black holes {\it stop} nuclear activity before less massive ones, since we describe the upper envelope of the BH mass distribution. In any case, this is indicative of anti-hierarchical `cosmic downsizing'
, which refers to the changing epoch of the peak of quasar activity for high and low mass black holes. 

\subsection{Comparison with previous results}
\begin{figure}
\centering
\includegraphics[width=0.5\textwidth, angle=0]{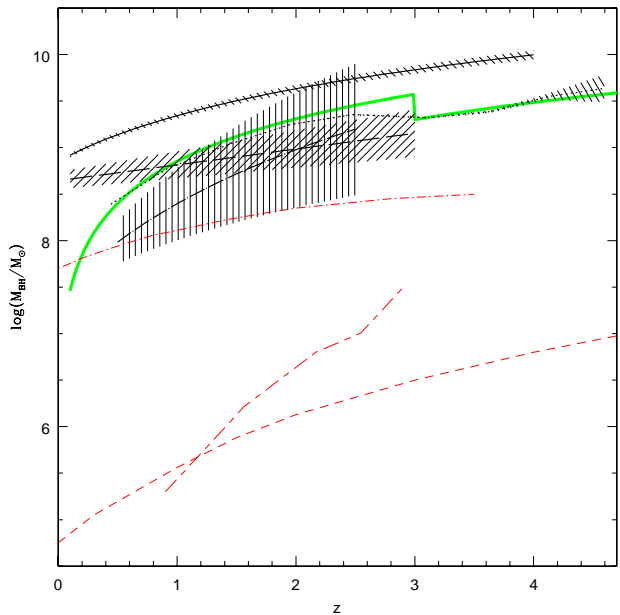}
\caption[]{ Thin solid line is the $M_{\rm BH}^{\rm max}(z)$ dependence studied in this work (Eq.~\ref{mmax_zall}). Dot-long dashed line is the best fit of $M_{\rm BH}(z)$ by Fine et al.~(2006) for a sample of 2dF quasars with luminosity around $L^*(z)$ at each redshift. Long dashed line is the best fit of $M_{\rm BH}$ as a function of redshift for a sample of $\sim100$ quasars with high quality spectroscopic observations (Decarli et al.~2009, in preparation). Shaded areas represent the uncertainties of the relationships. Dotted line shows the average $M_{\rm BH}$ as a function of redshift of the SDSS DR5 quasars (Shen et al.~2008) and the corresponding shaded area illustrates the error on the mean values per redshift bin. Green thick line represents the $M_{\rm BH}(z)$ trend corresponding to the minimum flux of the SDSS quasar selection algorithm (see text). 
Red lines are theoretical predictions from BH accretion models: dot-short dashed line is the average $M_{\rm BH}(z)$ by Merloni et al.~(2004), short-long dashed line represents $M_{\rm BH}(z)$ of BHs that accreted 5 per cent of their total mass (Marconi et al. 2004); short dashed line is the minimum BH mass required to exhibit nuclear activity as a function of $z$ (Shankar et al.~2009).}
\label{m_z_literature}
\end{figure}

In Fig.~\ref{m_z_literature} we present a global comparison between our results and the state-of-art knowledge on the redshift evolution of the quasar black hole mass. Our findings are consistent within the $1\sigma$ uncertainty with those by other authors (e.g.~Fine et al.~2006 and  Decarli et al.~in preparation, see Section 4.3 of Paper I for a detailed discussion). 
Our results are in agreement with recent theoretical predictions by e.g. Merloni (2004) and Merloni, Rudnick \& Di Matteo (2006), who studied the redshift evolution of the BH mass function and computed the average mass of supermassive active black holes as a function of $z$ through models accounting for radiatively efficient and inefficient BH accretion. 
There is also consistency with models by e.g.~Shankar, Weinberg \& Miralda-Escud\'e (2009), who studied AGN duty-cycles through 
BH accretion models to predict the $z$-dependence of the minimum BH mass at which supermassive black holes start to exhibit nuclear activity.  The results by Marconi et al.~(2004), who modeled with similar methods the $z$-dependence of the BH mass of active BHs that accreted a fixed fraction of their total mass (i.e., which entered their AGN phase), are also qualitatively consistent with our findings.

Fig.~\ref{m_z_literature} also shows the $M_{\rm BH}(z)$ trend corresponding to a minimum detected flux, which is representative of the effects of a Malmquist-type bias on the observed average BH mass redshift dependence. 
In detail, we considered a survey flux limit that roughly corresponds to that of the SDSS quasar selection algorithm ($i<19.1$ at $z<3$ and $i<20.2$ at $z>3$, see Section \ref{secprocedurec4} and references therein). From the luminosity dependence of the virial BH mass estimator (see Section 2) we estimate the corresponding mass limit at a given FWHM as a function of redshift. It is apparent from Fig.~\ref{m_z_literature} that the derived BH mass trend traces the observed average $M_{\rm BH}$ $z$-dependence of the SDSS quasars at any redshift, thus confirming that it is dominated by the Malmquist bias. On the other hand, our results deviate substantially from the predicted effects of the bias, particularly at low redshift.

\section{Summary and discussion}\label{sd}
We presented the application of a  procedure which allows the determination of the $z$-dependence of the BH mass to a large sample of quasars for which line-width and continuum measures are available. 

Combining the results from the \Hb, \MgII\ and \CIV\ samples of SDSS quasars, we found that the maximum mass of the active BH populations evolves as 
$(1+z)^{1.64\pm0.04}$
from $z\approx0.1$ to $z\approx4$. It is noticeable that the dependence on redshift of the typical mass of active black hole populations is substantially unchanged from $\frac{1}{10}$ of the actual age of the Universe to now, covering the region of maximum quasar activity ($z\sim2-3$). This is a clear manifestation of cosmic downsizing: the most massive black holes were more actively accreting early in the Universe and with time progressively less massive ones dominate the population of AGN. 

A comparison between the three emission lines studied in the work suggests some considerations on the corresponding BLR properties. 
Consistently with other independent indicators of the BLR geometry (considerations on the broad lines shape and from the geometrical factor $f$), we suggest that the \MgII\ BLR is essentially isotropic (or rather geometrically thick) and the Doppler velocities are practically purely rotational, as the assumptions produce a high quality description of the data distribution in the FWHM--luminosity plane. Consequently the \MgII\ line width is a good indicator of the virial velocity from which it is possible to infer the BH mass. On the other hand, there is increasing evidence that the  \CIV\ line emitting region is disc-like and hence this line is representative of the virial velocity only if the BLR is observed edge-on, i.e.~when the Doppler effect (which acts only on the velocity component parallel to the line of sight) is maximum.   The BH mass determinations from \CIV\ are then subject to an intrinsic error due to the uncertainty on the inclination angle and then on the virial velocity of the clouds. The \Hb\ line emitting region of the BLR is found to be closer to the isotropic case than \CIV\ one, but we suggest that for some objects the \Hb\ clouds are not at the equilibrium and are dominated by turbulent motion. This indicates that the \Hb\ emission line is a mediocre estimator of the BLR virial velocity and that the BH mass estimates from \Hb\ are subject to an uncertainty which is difficult to constrain. 

In conclusion, 
we suggest that an inflated disc broad-line region, in which the Carbon line is emitted in a flat inner disc while \MgII\ is produced in a geometrically thick outer region, can account for the observed differences in the FWHM distributions of the two emission lines. The \Hb\ broad emission line would be produced in clouds at the same distance from the central source as the \MgII\ ones, but the role of radiation pressure may be more relevant, accounting for non virial motion of the gas. We then
propose that the \MgII\ line and the corresponding 3000 \AA\ monochromatic luminosity are  preferable indicators of the BLR virial velocity and size respectively, and hence of the BH mass through Eq.~\ref{formulamassa}.

\section*{Acknowledgments}
We thank A.~Marconi for useful discussion and we are grateful to the anonymous referee for constructive criticism. We acknowledge support from ASI, INAF and the Italian Ministry of University. M.L.~acknowledges a fellowship from Insubria University.
\begin{footnotesize}

\end{footnotesize}
\label{lastpage}
\end{document}